%% file: ifacconf.tex
\documentclass{ifacconf}

\usepackage{longtable}
\usepackage{booktabs}
\usepackage{amsmath}
\usepackage{amssymb}

\usepackage{graphicx}      % include this line if your document contains figures
\usepackage{natbib}        % required for bibliography
\begin{document}
\begin{frontmatter}

\title{
% Combination of Quantitative Systems Pharmacology (QSP) and Physiology Based Pharmacokinetic Models (PBPK) for the Cholesterol Metabolism in the Human Body
A whole-body mathematical model of cholesterol metabolism and transport %\thanksref{footnoteinfo}
} 
% Title, preferably not more than 10 words.

\thanks[footnoteinfo]{Peter Emil Carstensen is at submission time funded by the The Novo Nordisk Foundation Center for Biosustainability, Technical University of Denmark, NNF20CC0035580.}

\author[DTU,NOVO]{Peter Emil Carstensen}
\author[DTU,NOVO]{Jacob Bendsen}
\author[DTU,NOVO]{Laura Hjort Blicher}
\author[NOVO]{Kim Kristensen}
\author[DTU]{John Bagterp Jørgensen}

\address[DTU]{Technical University of Denmark, Department of Applied Mathematics and Computer Science, DK-2800 Kgs. Lyngby, Denmark}
\address[NOVO]{Novo Nordisk A/S, DK-2880 Bagsværd, Denmark}

\input{tex/00Abstract}


\end{frontmatter}
%===============================================================================
\input{tex/01Introduction}

\input{tex/02CholesterolAndPhysiology}
% \input{tex/03 Model}
\input{tex/03aFlow}
\input{tex/03bDifferentialEquations}

\input{tex/03cReactions}
\input{tex/03dLiraglutide}
\input{tex/04Results}
\input{tex/05Discussion}

\input{tex/06Conclusion}

\bibliography{ifacconf}

\appendix

\input{tex/07Appendix}
\end{document}

%% file: tex/00Abstract.tex
\begin{abstract}                % Abstract of 50--100 words
% Cardiovascular diseases are the leading cause of death globally. Increased levels of plasma cholesterol are consistently associated with a heightened risk of cardiovascular disease. As a result, it is imperative that studies are conducted to determine the best course of action to reduce whole-body cholesterol levels. Mathematical models can provide direction on this. Therefore a whole-body mathematical model for the metabolism and transport of cholesterol, as well as a generalized approach for modeling metabolic networks with mass balances and gene regulation are proposed. This approach allows for efficiently modeling large metabolic networks in a systematic and compact way, utilizing ordinary differential equations and kinetic functions. The effect of lipid lowering drugs like statins work by inhibiting HMGCR and PCSK9 work by increasing the receptor recycling. The treatments were implemented and simulations showed similar results to what is reported in literature. Additionally, a sensitivity analysis of the current model shows, that the most effective drugs already target the parameters with the highest sensitivity. Mechanistic models serve as a valuable tool for designing therapeutic approaches aimed at regulating cholesterol levels and lipid compositions. The ability of the proposed model to simulate the impact of various drugs individually and in combination serve as a versatile platform for advancing personalized and effective interventions in the management of cardiovascular health.
Cardiovascular diseases are the leading cause of death. Increased levels of plasma cholesterol are consistently associated with an increased risk of cardiovascular disease. As a result, it is imperative that studies are conducted to determine the best course of action to reduce whole-body cholesterol levels. A whole-body mathematical model for cholesterol metabolism and transport is proposed. The model can simulate the effects of lipid-lowering drugs like statins and anti-PCSK9. The model is based on ordinary differential equations and kinetic functions. It has been validated against literature data. It offers a versatile platform for designing personalized interventions for cardiovascular health management.
\end{abstract}

%% file: tex/01Introduction.tex
\section{Introduction}

%First part
% Recent studies emphasize the significant impact of cholesterol levels on cardiovascular health, revealing a substantial 26\% reduction in major adverse cardiovascular events per 1 mmol/L reduction in low density lipoprotein cholesterol \citep{vallejo-vaz2018a}. 

The human body has several metabolic processes that continuously produces and uses cholesterol to maintain homeostasis. Cholesterol molecules are transported in lipoproteins and used in the cell for synthesis of hormones and vitamins as well as for cell membrane formation. Lipoproteins also have a vital function in supplying cells with free fatty acids for storage in the adipose tissue. Disruption in the lipoproteins homeostasis can lead to various diseases including atherosclerosis (cholesterol deposits in the blood vessel walls). Atherosclerosis can happen due to prolonged high cholesterol concentration in plasma.  This cholesterol formation may rupture and lead to a clot in the legs, the lungs, or the brain. This can cease blood flow and potentially be lethal \citep{miesfeld_mcevoy_2017}.

%Second part
The development and application of physiologically based pharmacokinetic (PBPK) models in chemical toxicology has grown steadily since their emergence in the 1980s. 
Whole-body mathematical models are useful to describe the biological interplay between cellular regulation and convective transport between organs through the blood stream. Such models allows for simulation of concentrations in organs, the transport rate across the membrane, and the cellular reactions \citep{paalvast2015a}.
Mathematical whole-body models allow for {\em in-silico} investigation of clinical trials prior {\em in-vivo} conduction of these trials. Potentially, this can accelerate the development of novel drugs and therapeutic intervention strategies.

%Third part
In the early 1900s, cholesterol was discovered to be a component in atherosclerosis. In the last couple of decades,  several computational models of cholesterol metabolism and transport have been proposed. \cite{paalvast2015a} evaluated computational models of  cholesterol metabolism in order to elucidate the regulation of intracellular cholesterol. \cite{Zhang:et:OverviewModels} reviewed several in-silico models of cholesterol metabolism and transport from the gene level to the overall transport in plasma. Most of the mathematical models reviewed could not provide the needed reliable integrated multi-scale description of organ metabolism and transport between organs. The models were often unable to reproduce experimental data and predict the outcome of a drug treatment \citep{paalvast2015a}. \cite{pratt2015a} formulated a model of hepatic lipid metabolism and \cite{toroghi2019a, sokolov2019a} considered mathematical models for cholesterol metabolism with drug induced effects from statins and anti-PCSK9.
%details of the  
% These models serve as a stepping stone for our model building, implemented with previous mathematical methods from \cite{Carstensen:Bendsen:etal:FOSBE:2022}

%Fourth part
A whole-body model of the human metabolism and transport of cholesterol, with the most relevant organs, herein a meal model and a pharmacokinetic model is proposed. The results of the simulations are compared to published clinical studies from \cite{peradze2019, matikainen2019a, aoki2020a, taskinen2021a, mok2023a}. The model provides a multi-scale mathematical model, that incorporates mass balances and genetic influences, through the use of ordinary differential equations and kinetic functions. Our model approach is systematic and compact, which enables efficient representation of large metabolic networks, as well as examination of the parameters most sensitive to changes. Our mathematical model serves as an in-silico tool that captures the dynamics of cholesterol metabolism under various metabolic and drug treatment conditions. \cite{Carstensen:Bendsen:etal:FOSBE:2022} provide the whole-body model building methodology, while \cite{Bendsen:Carstensen:MSc:2024} provide the specific details of the cholesterol metabolism model summarized in this paper. 

%Fifth part
The remaining part of this paper is structured as follows. Section \ref{sec:CholesterolTransport} describes the cholesterol metabolism and transport. Section \ref{sec:Model} presents the mathematical model and Section \ref{sec:Results} shows the simulation results. Section \ref{sec:Discussion} discusses the results and Section \ref{sec:Conclusion} briefly concludes.

%% file: tex/02CholesterolAndPhysiology.tex
\section{Cholesterol metabolism and transport}
\label{sec:CholesterolTransport}
\begin{figure}
    \centering
    \includegraphics[width = 0.95\columnwidth]{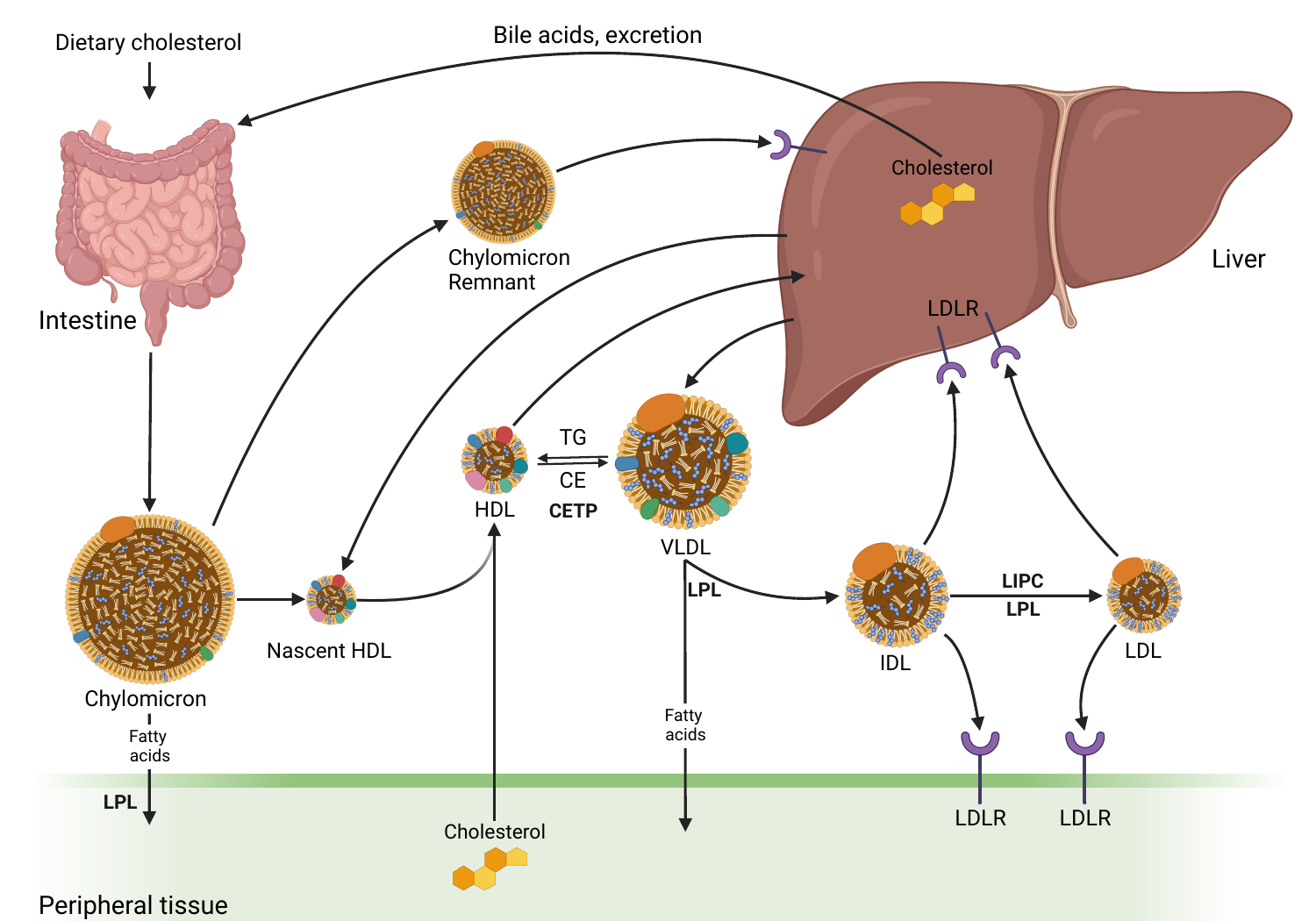}
    \caption{Lipoprotein metabolism. Triglyceride-rich particles are secreted from the intestine (Chylomicrons) or the liver (VLDL). Chylomicrons and VLDL particles undergo lipolysis in the circulation, which yields fatty acids to peripheral tissues. Remnants are taken up by the liver. IDL is taken up by the LDLR or further metabolized to LDL. HDL is formed in circulation from lipid-poor particles (nascent HDL), that are secreted from the liver or Chylomicrons. HDL carries cholesterol from peripheral tissues to the liver or exchanges cholesterol esters and triglycerides with lipoproteins in circulation. Adapted from \cite{lusis2008a} and created with BioRender.com.}
    \label{fig:cholesterol_transport}
\end{figure}

\begin{figure}
    \centering
    \includegraphics[width = 0.95\columnwidth]{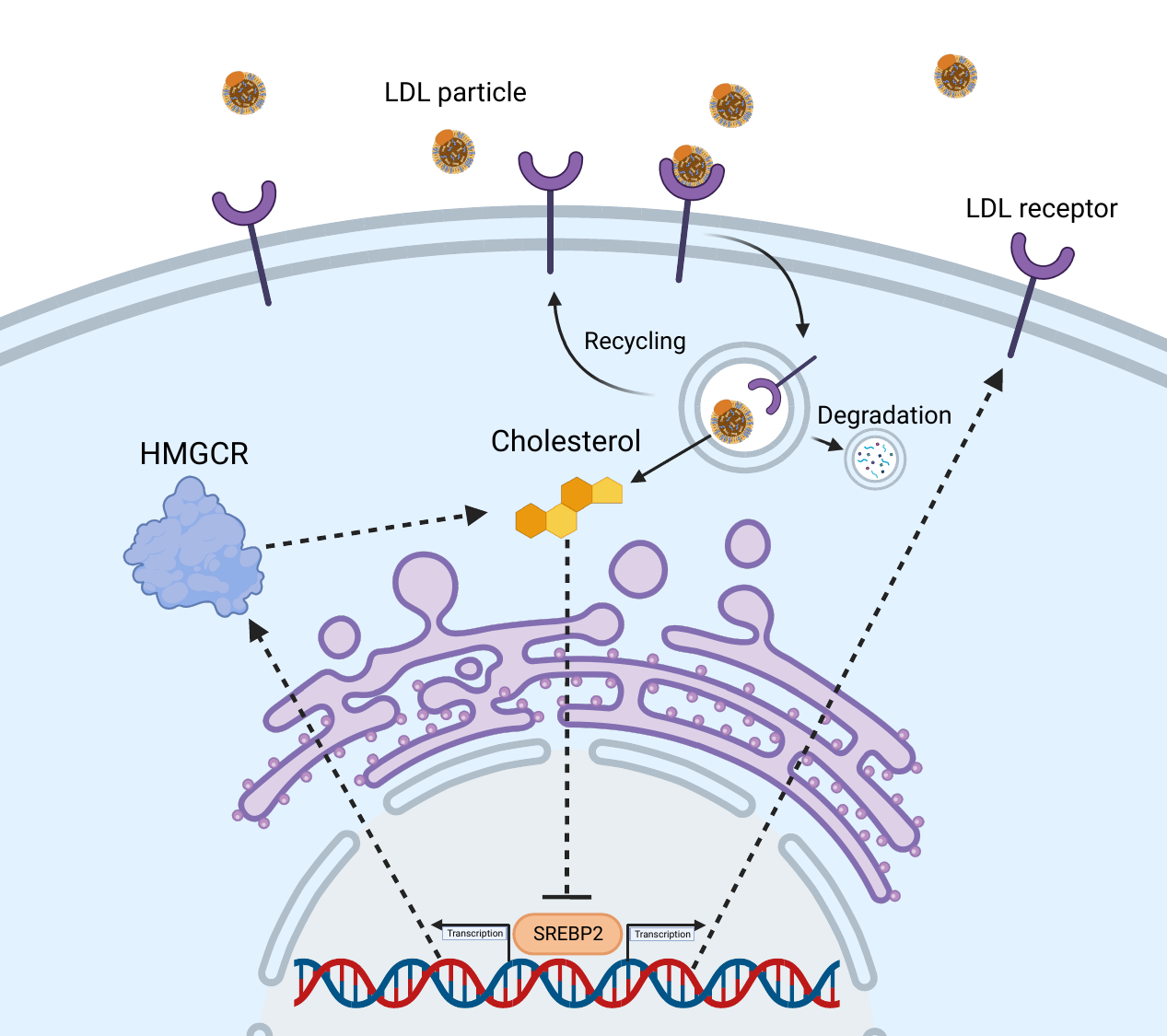}
    \caption{Cellular regulation of cholesterol and the LDLR. SREBP-2 binds to the HMGCR gene to produce HMGCR, that stimulates cholesterol synthesis. SREBP-2 also binds to the LDLR gene to produce LDLR. High cholesterol concentrations inhibit SREBP-2. LDL particles bind to LDLR in the interstitial space and the LDLR-LDL complex is internalized. If PCSK9 is bound to the LDLR, the receptor is degraded, otherwise it can be recycled to the surfaced. Created with BioRender.com.}
    \label{fig:Cholesterol_regulation}
\end{figure}

Figure \ref{fig:cholesterol_transport} shows the transport of cholesterol throughout the human body and the interplay of different lipoproteins (VLDL, IDL, LDL, Chylomicron, Remnant).
% Figure \ref{fig:cholesterol_transport} shows an overall diagram of how cholesterol is transported throughout the human body, and the interplay of the different lipoproteins. 
Cholesterol is transported in the body via lipoproteins. Dietary lipids are absorbed and packed into Chylomicrons. 
Chylomicrons consist of cholesterol, phospholipids, and free fatty acids (FFA). Lipoprotein lipases (LPL) release these FFA to peripheral tissues. 
% Chylomicrons are metabolized by lipoprotein lipases (LPL), that release fatty acids to tissues. 
The Chylomicron Remnants (Chylomicron without FFA) binds to receptors in the liver.
Non-dietary cholesterol is transported from the liver to tissues in very low density lipoproteins (VLDL). VLDL are metabolized by LPL into intermediate density lipoprotein (IDL), which can be further metabolized into low density lipoprotein (LDL). IDL and LDL binds to receptors in various tissues for degradation into cellular cholesterol.
High density lipoprotein (HDL) removes excess cholesterol from peripheral tissues. Through the cholesteryl ester transfer protein (CETP), HDL can also exchange triglycerides and cholesteryl esters with other lipoproteins.
Cholesterol is regulated at a cellular level by  sterol regulatory element-binding protein 2 (SREBP-2). SREBP-2 is inhibited by high cholesterol concentrations, as illustrated in Figure \ref{fig:Cholesterol_regulation} \citep{miesfeld_mcevoy_2017}.

Liraglutide is a GLP-1 receptor agonist, used for the treatment of type 2 diabetes, and chronic obesity (ref). Liraglutide also reduces the risk of cardiovascular disease (CVD) by reducing the cholesterol levels \citep{peradze2019, matikainen2019a, taskinen2021a}. According to \cite{peradze2019}, the cholesterol concentration levels are reduced with daily injections of 3.0 mg Liraglutide. This dose only has minor impact on cholesterol concentration in HDL, VLDL, and IDL. \cite{taskinen2021a} describe that Liraglutide can lead to reduced Chylomicron and VLDL concentrations. In turn they lead to decreased Remnant formation.
\begin{figure}
    \centering
    \includegraphics[width = \columnwidth]{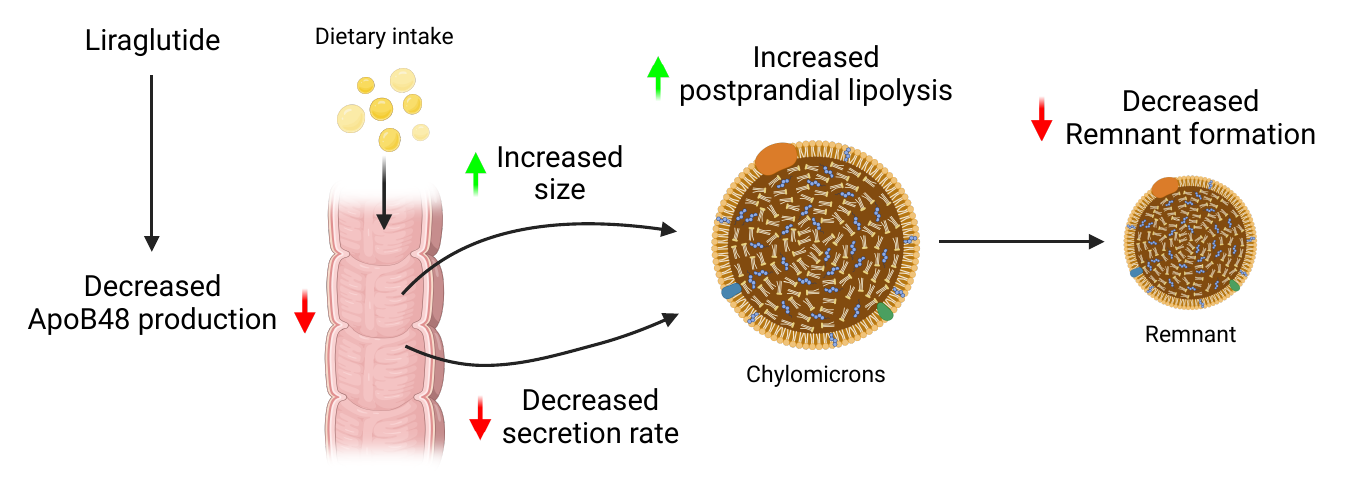}
    \caption{Proposed mechanism of action of Liraglutide on dietary intake and Chylomicrons. Liraglutide decreases (red arrow) ApoB48 production, leading to increased (green arrow) size of Chylomicrons and increased postprandial lipolysis. A decreased secretion due to fewer ApoB48 particles reduces the amount of Chylomicrons and leads to reduced Remnant formation. Adapted from \cite{taskinen2021a} and created with BioRender.}
    \label{fig:Proposed_mechanism_Liraglutide}
\end{figure}
Figure \ref{fig:Proposed_mechanism_Liraglutide} shows the proposed mechanism of Liraglutide on dietary cholesterol intake \citep{taskinen2021a}. Liraglutide has a direct suppressive action on ApoB48 synthesis in the gut. Liraglutide reduces the postprandial production rate of Chylomicron-ApoB48 by 60\%. As a consequence of the reduced ApoB48 availability, the Chylomicrons assemble themselves in larger particles. The larger Chylomicron particles can be lipolysed more rapidly than the smaller Chylomicron particles and have a reduced direct clearance. 
\cite{taskinen2021a} also compare blood samples before and after treatment with Liraglutide. They observe a smaller concentration of ApoB48 containing particles in treatments with Liraglutide. This is also supported by previous experiments \citep{matikainen2019a}.
% , for those patients treated with Liraglutide compared to the baseline. 
Additionally, the reported data in \cite{taskinen2021a} show a reduction in the accumulation of liver fat. The VLDL secretion rate scales with the amount of liver fat. Therefore, less liver fat means less VLDL secretion \citep{taskinen2021a}. %less liver fat reduces the secretion rate of VLDL.

% triglycerides in VLDL. 
% \cite{taskinen2021a} concludes by saying that the overall effect of Liraglutide treatment for reduction in cholesterol, is to diminish the generation of remnant lipoproteins, such that there is less overall ApoB48 containing particles. 

%% file: tex/03aFlow.tex
\section{Model}
\label{sec:Model}
\begin{figure}
    \centering
    \includegraphics[width = 0.4\columnwidth]{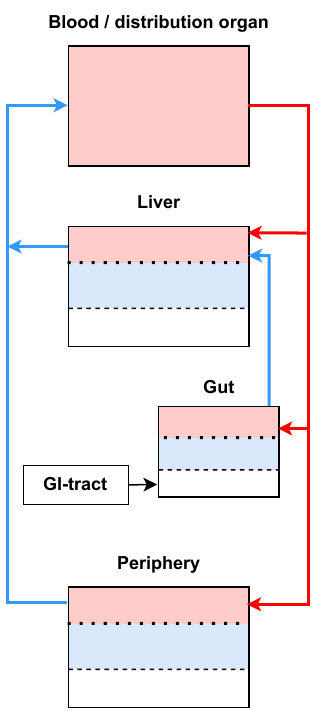}
    \caption{The cholesterol flow model.
    % The flow model with the most important organs in relation to cholesterol metabolism. 
    The Periphery should be perceived as a lumped compartment. 
    Dietary cholesterol is absorbed in the gut cellular space.
    % Arterial and venous blood are not modelled explicitly.
    }
    \label{fig:Flow_model}
\end{figure}

The model in this paper is a connected system with four organs: Blood (B), Liver (L), Gut (G), and Periphery (P). Periphery is a lumped compartment of all other organs. Figure \ref{fig:Flow_model} shows the connected system.  The GI-tract refers to a meal uptake model \citep{Ritschel:etal:MealModels:2023}. Figure \ref{fig:Organ_model} illustrates the additional division of organs into Vascular, Interstitial and Cellular. Figure \ref{fig:Organ_model} is denoted the organ model. %Additionally, the blood compartment is not further divided in three sub compartments, as its main function is to distribute metabolites through its flow. 

% \begin{table}
% \centering
% \caption{Blood flow rates as L/min and \% of cardiac output for a 73 kg European Caucasian male, and volumes of each included compartment from \cite{Valentin:ICRP:2002}. The volume for the blood compartment consists of the venous blood, arterial blood and lung. The liver is only the liver. The gut is the large intestine, pancreas, small intestine, spleen, stomach and portal vein. The periphery tissue is the bone, brain, fat, gonads, kidney, muscle, skin and heart.}
%  \label{tab:PKsim_ref}
%  \begin{tabular}{lll}
%  % \toprule
%   & Q [L/min] (\%) & V [L]  \\ %\\
%  \toprule
% Blood     & 6.09 (100)   &  2.59 \\
% Liver     & 0.43 (7.1)   &  2.38 \\
% Gut       & 1.22 (20.0)  &  2.74 \\
% Periphery & 4.44 (72.9)  &  65.29\\ 
% \bottomrule
% \end{tabular}
% \end{table}

%% file: tex/03bDifferentialEquations.tex
% \subsection{Equations}
Consider the set of organs, 
$\mathcal{O} = $ $\{$B, L, G, P$\}$. Then the mass balances for the vascular part is
\begin{subequations}
\label{eq:overall}
\begin{alignat}{3}
    \label{eq:DC_v}
    V_{Ov} \dot{C}_{Ov} &= q_{O} - C_{Ov} Q_{O,out} + f_{Ov} \qquad && \forall O \in \mathcal{O} \\
    q_{O} &= \sum_{\bar O \in \mathcal{I}_O} C_{\bar Ov} Q_{\bar O} \qquad && \forall O \in \mathcal{O} \\
    Q_{O,out} &= \sum_{\bar O \in \mathcal{I}_O} Q_{\bar O} && \forall O \in \mathcal{O} 
\end{alignat}
\end{subequations}
The mass balances for the interstitial organ spaces and the cellular organ spaces can be described as
\begin{subequations}
\begin{alignat}{3}
    \label{eq:DC_i}
    V_{Oi} \dot{C}_{Oi} &= f_{Oi} \qquad && \forall O \in \mathcal{O} \setminus \{B\} \\
    \label{eq:DC_c}
    V_{Oc} \dot{C}_{Oc} &= f_{Oc} && \forall O \in \mathcal{O} \setminus \{B\}
\end{alignat}
\end{subequations}
$V_{Ov}$, $V_{Oi}$ and $V_{Oc}$ the vascular, interstitial and cellular volumes respectively. $\mathcal{I}_O$ is the set of inflows to organ $O$. 
$f_{Ov}$, $f_{Oi}$ and $f_{Oc}$ are the vascular, interstitial and cellular transport and reaction rates and can be expressed as
% The subscript $v$ indicates the vascular space, $i$ the interstitial space, and $c$ the intracellular space.
\begin{subequations}
\label{eq:metabolic}
\begin{alignat}{3}
f_{Ov} &= -T_{Ovi} V_{Ov} + T_{Oiv} V_{Oi} + R_{Ov} V_{Ov}  \\
\begin{split}
f_{Oi} &= T_{Ovi} V_{Ov} - T_{Oiv} V_{Oi} \\ & \quad - T_{Oic} V_{Oi} + T_{Oci} V_{Oc} + R_{Oi} V_{Oi}
\end{split} \\
f_{Oc} &= T_{Oic} V_{Oi} - T_{Oci} V_{Oc} + R_{Oc} V_{Oc}  + s_{Oc}
\end{alignat}
\end{subequations}
Transport between compartments occurs through a concentration gradient and is expressed as
\begin{subequations}
\label{eq:OrganTransportRateModelGeneral}
\begin{alignat}{3}
    T_{Ovi} &= k_{Ovi}C_{Ov}  \qquad && \forall O \in \mathcal{O} \setminus \{B\}\\
    T_{Oiv} &= k_{Oiv}C_{Oi}  \qquad && \forall O \in \mathcal{O} \setminus \{B\}\\
    T_{Oic} &= k_{Oic}C_{Oi}  \qquad && \forall O \in \mathcal{O} \setminus \{B\}\\
    T_{Oci} &= k_{Oci}C_{Oc}  \qquad && \forall O \in \mathcal{O} \setminus \{B\}
\end{alignat}
\end{subequations}
% Except for those reactions that make use of receptor-mediated uptake, in which the transport reactions are governed by,
% \begin{equation}
%     T_R = \alpha + \frac{C_{fR}}{\sum C_{R}} (\alpha - 1) \\
% \end{equation}
% $C_{fR}$ is the concentration of free receptors and $C_R$ is the concentration of bound receptors, to e.g. LDL, IDL or remnants, and $\alpha$ is the percentage of receptors that get recycled. 
% A diagram of the organ model is shown in Figure \ref{fig:Organ_model}.
The production rates in each compartment
\begin{subequations}
\label{eq:ChemicalModel}
\begin{alignat}{3}
R_{Ov} &= \nu_{Ov}' r_{Ov}(C_{Ov}) \qquad && \forall O \in \mathcal{O} \\
R_{Oi} &= \nu_{Oi}' r_{Oi}(C_{Oi}) \qquad && \forall O \in \mathcal{O} \setminus \{B\} \\
R_{Oc} &= \nu_{Oc}' r_{Oc}(C_{Oc}) \qquad && \forall O \in \mathcal{O} \setminus \{B\}
\end{alignat}
\end{subequations}
are obtained from the stoichiometry, $\nu$, and reaction rates, $r(C)$, whose kinetics depend on the current concentration of the species for each compartment in all organs $\mathcal{O}$. 

% Those reactions that make use of receptor-mediated uptake, could be expressed as Michaelis-Menten kinetics. 
The term $s_{Oc}$ represents a sink or source function within the cellular space. A source could provide the intestines with dietary cholesterol and a sink could be the conversion of cholesterol to bile in the liver. 
% A meal model from \cite{Ritschel:etal:MealModels:2023} is used as a source.
\begin{figure}
    \centering
    \includegraphics[width = \columnwidth]{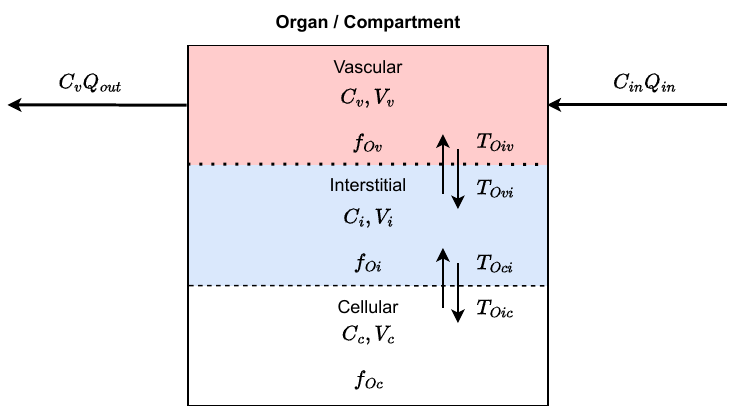}
    \caption{The organ model with 3 compartments: Vascular, Interstitial, Cellular. The metabolism in each compartment is described by $f_{O}$, the transport between compartments is defined by $T_{O}$ and the blood flow between organs is defined by $C\cdot Q$. 
    % The organ is split into three compartments, from top to bottom: vascular, interstitial space, intracellular space.
    }
    \label{fig:Organ_model}
\end{figure}

%% file: tex/03cReactions.tex
\subsection{Reactions}
\begin{figure}
    \centering
    \includegraphics[width = \columnwidth]{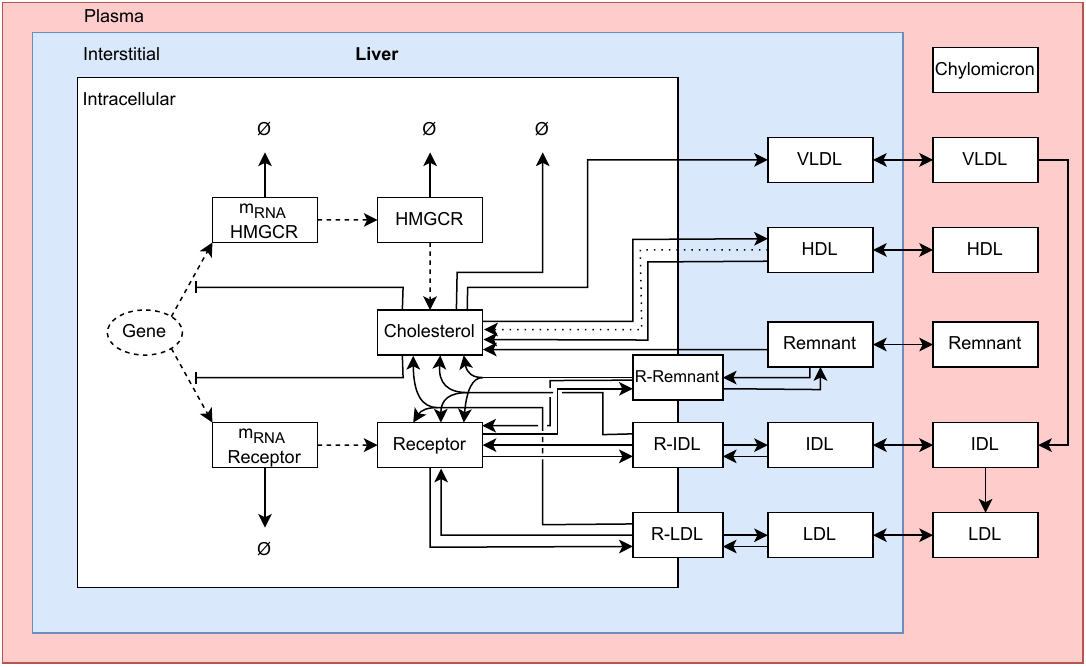}
    \caption{The cholesterol metabolism in the liver. The white area is the intracellular space, the blue area is the interstitial space, and the red area is the vascular space. The solid arrows represent mass transfer, the dashed arrows represent activation, the dotted arrows represent transfer of cholesterol molecules, the blunt arrows represent inhibition, the bidirectional arrows represent diffusion, and the dashed circle represents the nucleus of the cell.}
    \label{fig:Liver_metabolism}
\end{figure}

Figure \ref{fig:Liver_metabolism} shows the reactions and transport that occur within the liver. VLDL and Remnant particle transport only occurs in the liver. The biochemical reaction model for cholesterol is the same in the Liver, the Gut, and the Periphery. The transport varies from organ to organ, where the transport is greater in the Liver. The Gut secretes dietary Chylomicrons and the Periphery secretes excess cellular cholesterol to circulating HDL particles. 

%% file: tex/03dLiraglutide.tex
\subsection{Liraglutide pharmacokinetic-pharmacodynamic model}
The pharmacokinetics of Liraglutide is simulated by a sligthly modified model in \cite{watson2010a}, i.e. a one compartment model with first order absorption and first order elimination.
% The pharmacokinetics of Liraglutide is modelled by a one compartment model with first order absorption and first order elimination
\begin{subequations}
\begin{align}
     \dot{A}_1 &= u(t) - K_A A_1 \\
     V\dot{D}_1 &= K_A A_1 - CL \ D_1
\end{align}
\end{subequations}
$u(t)$ is the dosing rate of Liraglutide, $K_A$ is the absorption constant, $A_1$ is the amount of Liraglutide in the absorption compartment. $V$ is the distribution volume of Liraglutide, $CL$ is the clearance, and $D_1$ is the concentration of Liraglutide in the distribution compartment \citep{watson2010a}. 
The pharmacodynamics, i.e. the effect, of Liraglutide ($D_1$) on Chylomicrons is implemented as activation of lipolysis and inhibition of secretion. It is assumed that all organs has the Liraglutide concentration $D_1$.

% A PK model of Liraglutide is simulated alongside the whole-body model. 

% A two compartment model with first order absorption and first order elimination, can be described by,

% $K_A$ is the absorption constant in $1/h$, $A_1$ is the absorption compartment in $\mu g$. $V$ is the distribution volume of Liraglutide in $L$, $CL$ is the clearance in $L/h$ and $D_1$ is the distribution compartment in $\mu g/L$  \citep{watson2010a}. 
% Table \ref{tab:Watson} contains the parameters used in the PK model.

% \begin{table}
% \centering
%  \caption{Parameter values used in the PK model proposed by Watson et al. \citep{watson2010a}. The volume of distribution, clearance and dose are body weight specific. The volume and clearance also depend on the bioavailability of Liraglutide.} 
% \label{tab:Watson}
% \begin{tabular}{llll}
%  % \toprule
%   Description & Parameter & Value & Unit  \\ %\\
%  \toprule
% Body weight         & BW    & 73                         & Kg      \\
% Bioavailability     & F     & 0.51                       & -       \\
% Volume              & V     & 0.17 $\cdot$ BW $\cdot$ F  & L/kg    \\
% Clearance rate      & CL    & 0.012 $\cdot$ BW $\cdot$ F & L/h/kg  \\
% Absorption constant & K$_A$ & 0.104                      & 1/h     \\
% Dose                & D     & [0.6, 1.2, 1.8]            & mg      \\
% \bottomrule
% \end{tabular}
% \end{table}

%% file: tex/04Results.tex
\section{Results}
\label{sec:Results}
\subsection{Liraglutide}
\begin{figure}
     \centering
     \includegraphics[width=\columnwidth]{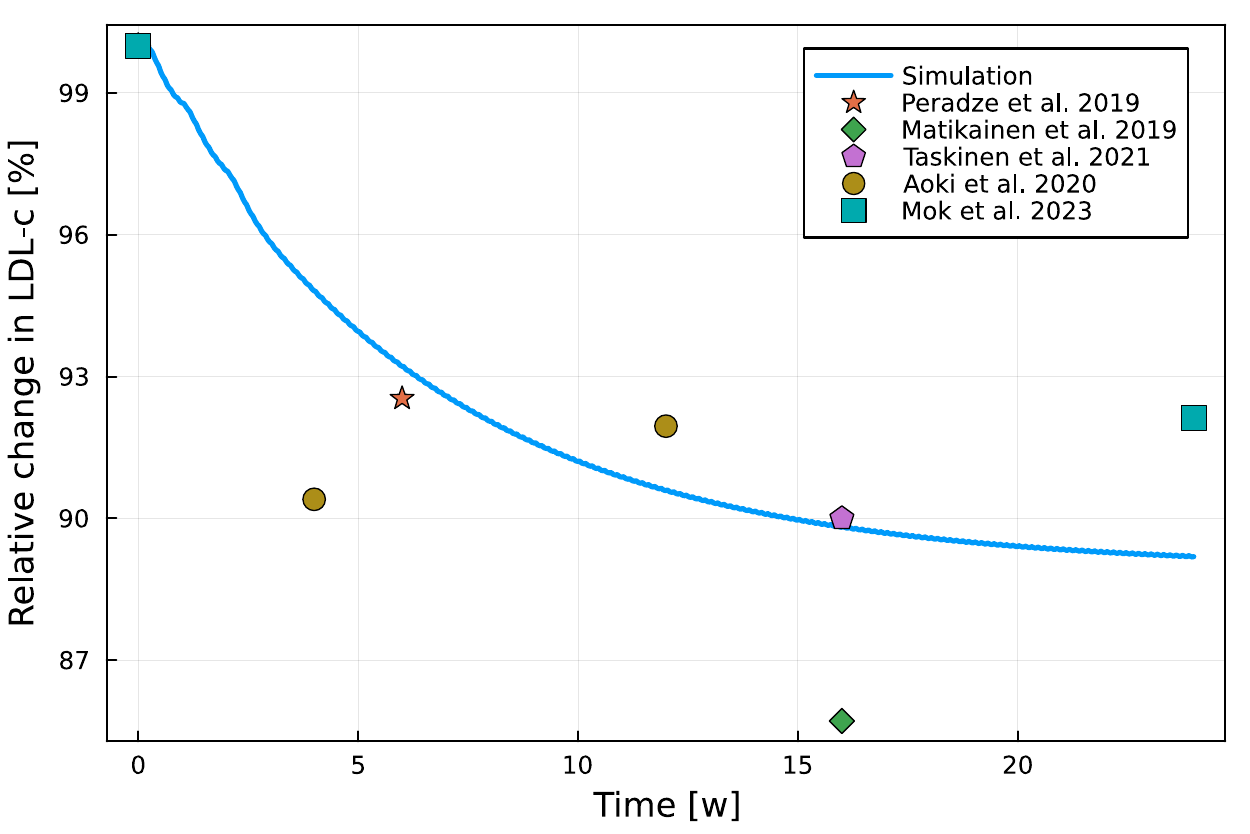}
     \caption{Relative drop in LDL-c after 24 weeks treatment with incremental increases in dose until 1.8 mg/day Liraglutide. The data points are digitized data.}
     \label{fig:100kg_relative}
\end{figure}

Liraglutide dosing is simulated as an incremental dosing regimen. The doses are: 0.6 mg/day in week 1; 1.2 mg/day in week 2; 1.8 mg/day in the subsequent weeks. Figure \ref{fig:100kg_relative} shows a simulated 24 week Liraglutide therapy. The simulated relative reduction compares well with experimental data from the literature \citep{peradze2019, matikainen2019a, aoki2020a, taskinen2021a, mok2023a}. The simulated therapy results in a 10\% LDL cholesterol (LDL-c) reduction after 24 weeks.
% simulation of relative reduction in LDL cholesterol (LDL-c) is simulated with 24 weeks of Liraglutide treatment. The rate of reduction and the final total reduction of around 10\% is seen to be in agreement with the digitized data.

\subsection{Statins}
\begin{figure}
    \centering
    \includegraphics[width = \columnwidth]{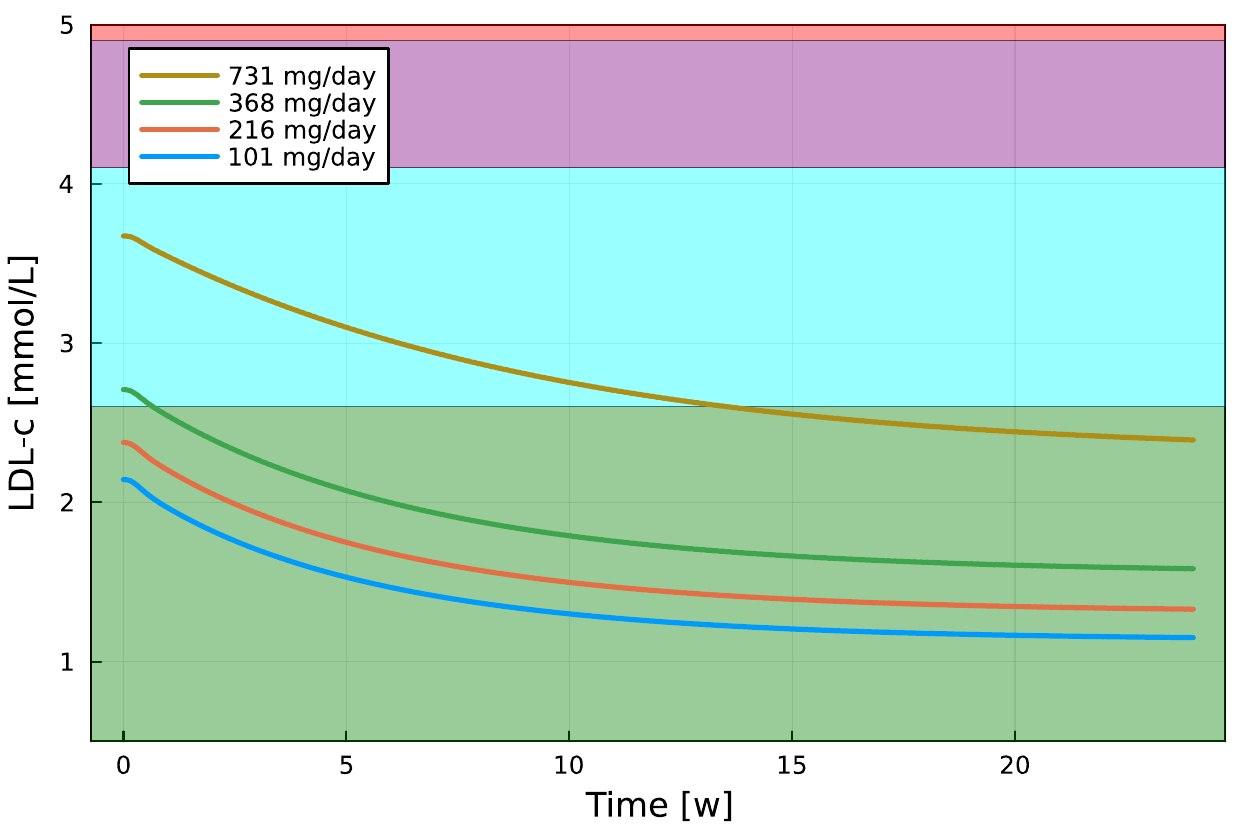}
    \caption{The simulated LDL-c for persons in statin treatment over 24 weeks with different diets (cholesterol intake per day).
    % Simulation results of the LDL-c for different diets on statin treatment over 24 weeks. 
    Red: Very high, Purple: High, Cyan: Borderline high, and Green: Optimal \citep{Panel:CholesterolLevels}.
    }
    \label{fig:statin_24_week_diff_diet}
\end{figure}
Figure \ref{fig:statin_24_week_diff_diet} shows simulations of statin therapy for different diets. 
The effect of statins are simulated by inhibition of the endogenous production of cholesterol by 85\% \citep{paalvast2015a}. 
Figure \ref{fig:statin_24_week_diff_diet} show a simulation of LDL-c reduction in a person treated with a statin therapy over 24 weeks combined with different dietary cholesterol intake.
% The largest simulated dietary cholesterol intake results in an LDL-c reduction of 35\%, whereas the smallest simulated intake results in a steady-state reduction of 47\%. 
% Consequently, statin treatment works in all cases but is most efficient with low dietary cholesterol intake.

\subsection{Anti-PCSK9}
\begin{figure}
    \centering
    \includegraphics[width = \columnwidth]{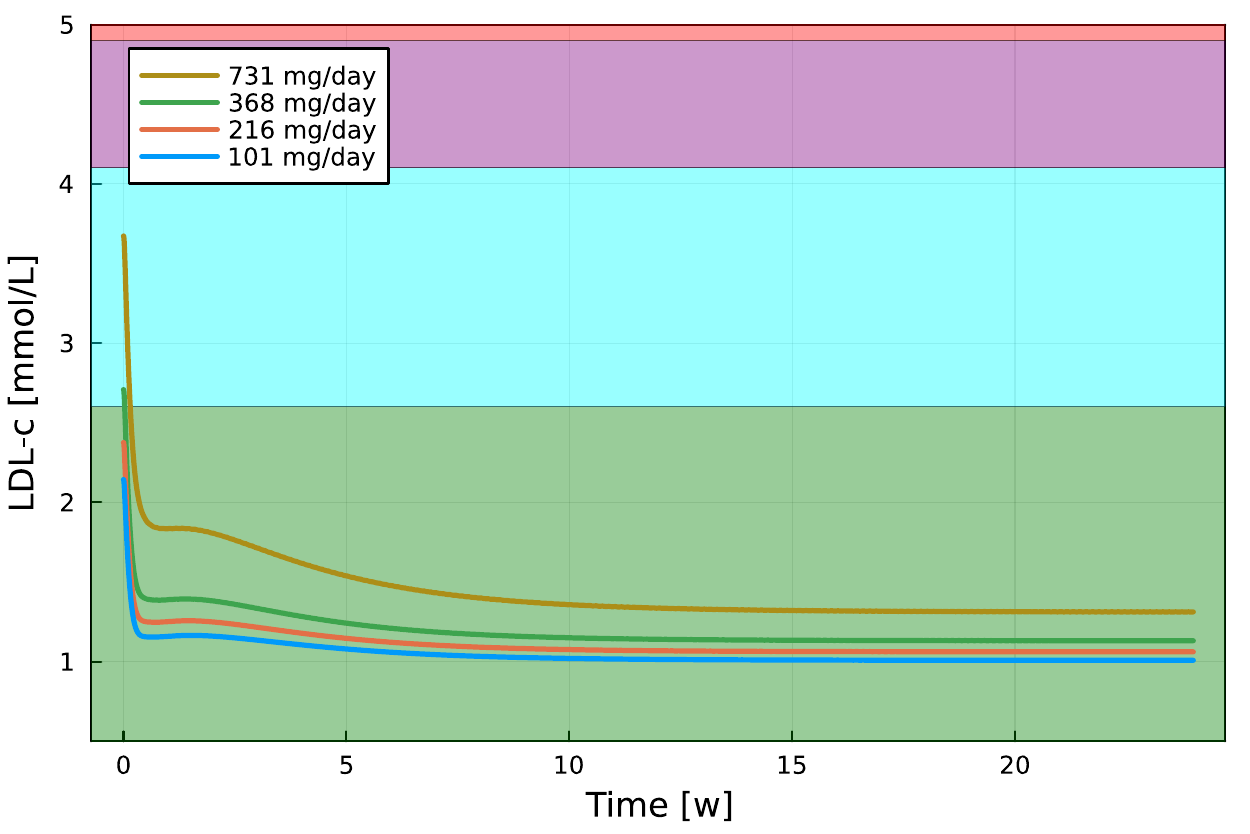}
    \caption{The simulated LDL-c for persons in anti-PCSK9 treatment over 24 weeks with different diets (cholesterol intake per day).
    % Simulation results of the LDL-c for different diets on anti-PCSK9 treatment over 24 weeks. 
    Red: Very high, Purple: High, Cyan: Borderline high, and Green: Optimal  \citep{Panel:CholesterolLevels}.
    }
    \label{fig:anti_pcsk9_24_week_diff_diet}
\end{figure}
Figure \ref{fig:anti_pcsk9_24_week_diff_diet} shows simulations based on anti-PCSK9 therapy over 24 weeks. The effect of anti-PCSK9 is modelled as a decreased receptor degradation \citep{Bendsen:Carstensen:MSc:2024}.
% an increased rate of receptor-recycling \citep{Bendsen:Carstensen:MSc:2024}. %The recycle parameter is increased from 0.7 to 0.95. %70\% to 95\%. 
%
%results in an immediate effect from the increased amount of receptors available. It can be seen, 
%
The after-treatment steady-states vary according to dietary cholesterol intakes. The largest simulated cholesterol intake results in a LDL-c steady-state of 1.31 mmol/L (64\% reduction). The smallest simulated dietary cholesterol intake results in a steady-state of 1.01 mmol/L (53\% reduction). Hence, in anti-PCSK9 treatment, the dietary intake of cholesterol on the final LDL-c level is small.

\subsection{Statins and anti-PCSK9}
\begin{figure}
    \centering
    \includegraphics[width = \columnwidth]{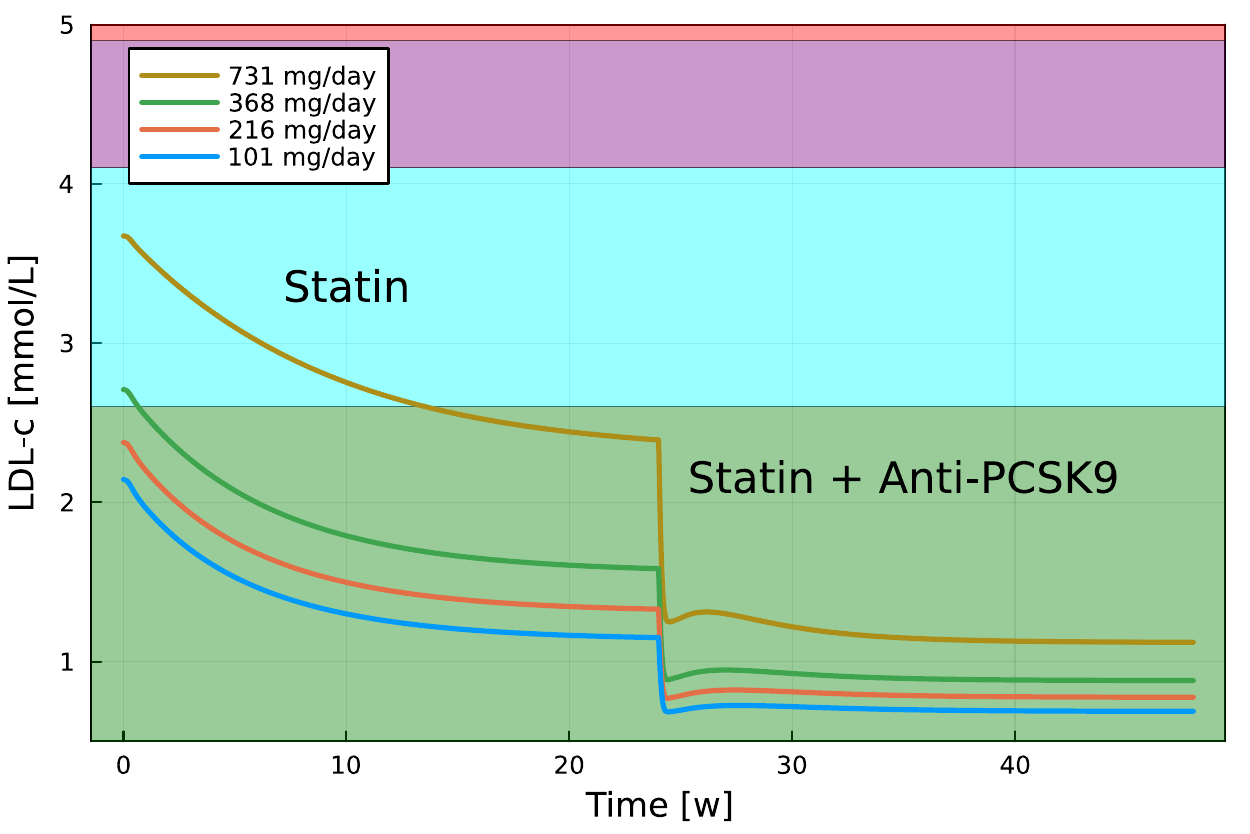}
    \caption{Simulated LDL-c for different diets (cholesterol intakes) in a therapy with 24 weeks of statin dosing followed by 24 weeks of combined statin and anti-PCSK9 dosing. Red: Very high, Purple: High, Cyan: Borderline high, and Green: Optimal \citep{Panel:CholesterolLevels}.}
    \label{fig:statin_anti_pcsk9_24_week_diff_diet}
\end{figure}

In a study by \cite{mckenney2012a}, the effects of anti-PCSK9 was investigated in patients who had already been on statin treatment for a period greater than six weeks. \cite{mckenney2012a} show that the LDL-c could be further reduced by 40-72\%. This is simulated by a 24 week period of treatment with statins and then additional 24 weeks of statin and anti-PCSK9 treatment. Figure \ref{fig:statin_anti_pcsk9_24_week_diff_diet} shows simulated results similar to the results by \cite{mckenney2012a}.

% The initial 24 weeks follow the same dynamics as seen in the statin simulations in \ref{fig:statin_24_week_diff_diet}, as well as the same initial reduction in LDL-c. As anti-PCSK9 is introduced as an additional drug, the LDL-c reduces by an additional 63\% for the largest dietary intake, and for the smallest dietary intake an additional reduction of 41\% can be seen. Showing the model is in agreement with published reduction intervals. 

\subsection{Sensitivity analysis}
\label{sec:SensitivityAnalysis}
\begin{figure}
    \centering
    \includegraphics[width = \columnwidth]{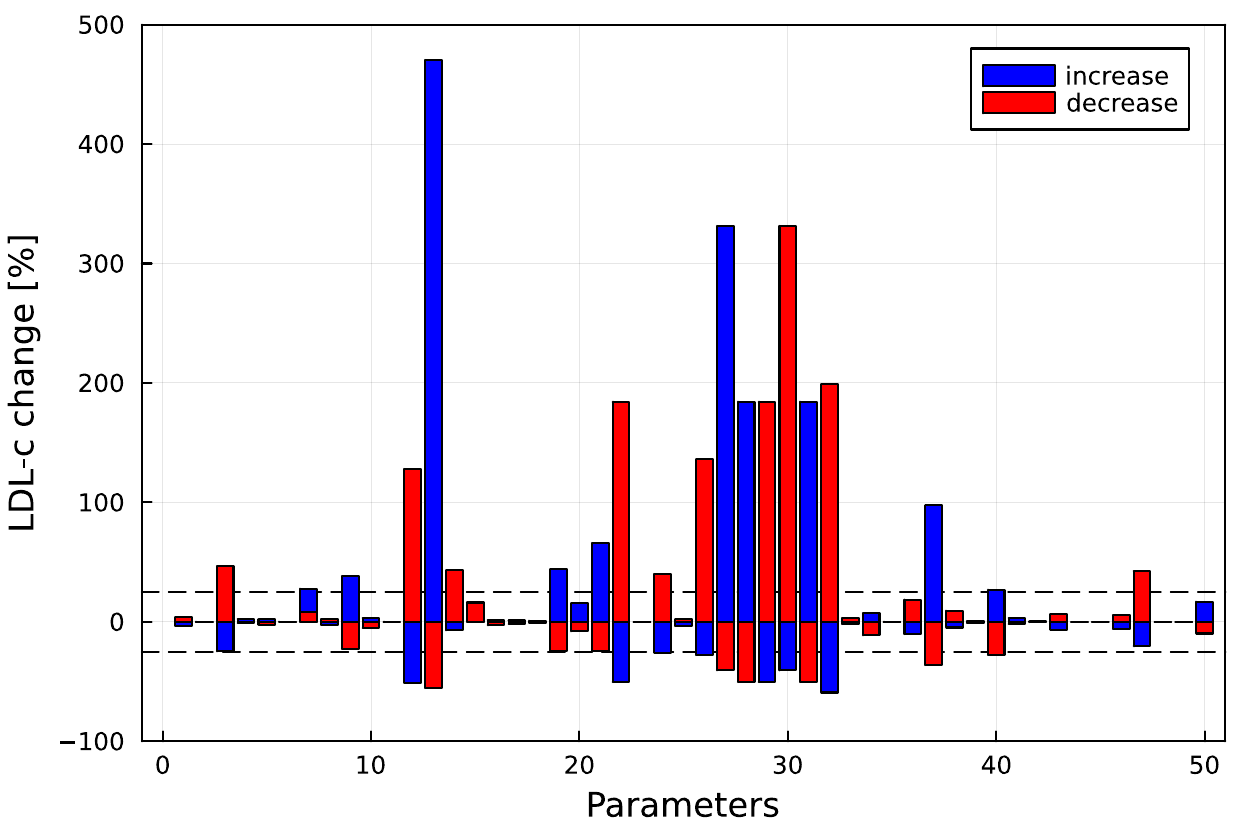}
    \caption{Bar plot of the result of the global sensitivity analysis. Doubling parameter value: \textit{increase} (blue), halving parameter value: \textit{decrease} (red). The dashed lines indicate the cutoff of $\pm 25\%$ change in LDL-c concentrations at steady state.}
    \label{fig:100kg_sensitivity_bar}
\end{figure}

A sensitivity analysis is performed for all parameters related to metabolism and transport. This is done by separately doubling or halving all parameter values in relation to their nominal value and simulating of the model for 52 weeks. Figure \ref{fig:100kg_sensitivity_bar} shows the effect on LDL-c for each of the 50 parameters in the model.
Twenty parameters are identified to have a significant effect on the LDL-c concentrations. The parameter $k_{VLDL\ sec}$ (parameter 13) increases the total amount of lipoproteins in the bloodstream and leads to the highest increase in LDL-c. The parameter $\delta_c$ (parameter 32) is related to the degradation rate of cholesterol in the liver. This parameter leads to the largest reduction in the LDL-c concentration. 
The model is sensitive to the biosynthesis of cholesterol, as 5 of the 20 most significant parameters are associated with the endogenous production of cholesterol.
% The model is sensitive to the biosynthesis of cholesterol as 5 parameters associated with the biosynthesis are among the 20 parameters to which the model is most sensitive. 
%
%The analysis additionally shows that the model is sensitive to parameters associated with biosynthesis of cholesterol, as it accounts for five of the 20 most sensitive parameters. 
%
% This is a natural part of the current model, as a higher production will lead to more cholesterol, which in turn will accumulate as LDL. \\
% There are three sensitive parameters associated with LDL receptor production, as this will affect the amount of LDL taken up by the tissues.  
%
The parameter associated with receptor-recycling has a high influence on the LDL-c concentration.  
% It has been varied less than the other parameters in the sensitivity analysis, because it has an upper limit. 
Increasing this parameter mimics the effect of anti-PCSK9.

% The parameter $\alpha$, corresponds to the percentage of receptors that are recycled and has a high influence. It has been varied less than the other parameters in the sensitivity analysis, because it has an upper limit. Increasing this parameter mimics the effect of anti-PCSK9.

%% file: tex/05Discussion.tex
\section{Discussion}
\label{sec:Discussion}
% TODO: PETER UPDATES

Statins and anti-PCSK9 impact different parameters within the model. Simulations shows that the proposed model is capable of incorporating these drugs. Statins vary in strength for their reduction of cholesterol. Statins can provide 20-50\% reduction in LDL-c \citep{scharnagl2001a}. The simulated statin therapy showed a reduction of 35-47\% for the four different cholesterol intakes. Anti-PCSK9 therapy reduces cholesterol levels from 40-70\% \citep{lambert2012a}. The simulated anti-PCSK9 treatment shows a 53-64\% reduction in LDL-c. The model's analysis of therapeutic interventions with the combination of statins and anti-PCSK9 treatments, shows an additional 41-63\% reduction in LDL-c. This is also within the 40-72\% additional reduction reported by \cite{mckenney2012a}. 

% Given that reductions in LDL-c were found to be associated with large reductions in the relative risk for MACE, the results highlight the potential for reductions in the annual number of deaths through therapeutic drug interventions.
The sensitivity analysis shows that most of the significant parameters are already targets of current lipid lowering drugs. Two new possible targets have been identified. If a drug can inhibit the secretion rate of VLDL, our model shows a significant impact on the LDL-c concentration. If a drug can increase the degradation rate of cholesterol then our model also shows significant reduction in LDL-c. Inhibition of VLDL secretion will inevitably increase the cellular cholesterol concentration. Given an increased cellular cholesterol, the degradation rate of cholesterol becomes highly relevant. A drug that targets both, would be effective in lowering the lipid composition. Further validation of the model is required to determine if this is a valid drug target.

%% file: tex/06Conclusion.tex
\section{Conclusion}
\label{sec:Conclusion}
We provide a whole-body mathematical model of the cholesterol metabolism in man. The model can be used to simulate LDL-c lowering therapies with Liraglutide, statins, anti-PCSK9 and combinations of statins and anti-PCSK9. A model sensitivity analysis demonstrates that the biosynthesis of cholesterol is important. The model and the simulations indicate that {\em in-silico} clinical test may be a valuable supplement to guide {\em in-vivo} clinical trials. In addition, the model and the modeling activity provide an advanced  understanding of cholesterol metabolism and transport. The corresponding insights enable further research and refinement of mathematical models for dynamical simulation of cholesterol metabolism and possible therapies.  Future work may extend the model with individual population metrics (allometric scaling) to design personalized intervention strategies for cardiovascular health.

% in-silico models to guide future therapeutic developments and interventions.